\documentclass[
 reprint,
 amsmath,amssymb,
 aps,
 prb,
]{revtex4-1}

\usepackage{dcolumn}% Align table columns on decimal point
\usepackage{bm}% bold math
\usepackage[pdftex]{graphicx,color}
\usepackage{comment}
\usepackage{natbib}
\usepackage{here}
\usepackage[normalem]{ulem}
\usepackage{hyperref}% add hypertext capabilities
\usepackage{braket}
\usepackage{url}

\newcommand{\red}[1]{\textcolor{red}{#1}}

\begin{document}

\preprint{APS/123-QED}

\title{
Channel-selective non-Fermi liquid behavior in 
the two-channel Kondo lattice model under a magnetic field
}

\author{Koji Inui and Yukitoshi Motome}
\affiliation{
 Department of Applied Physics, The University of Tokyo, Hongo, Tokyo 113-8656, Japan
}

\begin{abstract}
Stimulated by anomalous behaviors found in non-Kramers $f$-electron systems in an applied magnetic field, we study a two-channel Kondo lattice model by using 
a cluster extension of the dynamical mean-field theory combined with the continuous-time quantum Monte Carlo method. 
We include the effect of the external magnetic field in two ways: 
the Zeeman coupling to conduction electron spins and an effective coupling to the quadrupole degree of freedom through the crystalline electric field splitting. 
We show that the magnetic field suppresses the antiferroic-spin order (physically, corresponding to the antiferroic-quadrupole order), 
and yields a channel-selective non-Fermi liquid state 
where one of the two channels (physically, spin-up or -down) exhibits 
non-Fermi liquid behavior while the other shows Fermi liquid behavior, before entering the Fermi liquid regime 
in higher fields. 
This anomalous state appears in a dome-shaped region which extends from inside the antiferroic-spin
ordered phase to the paramagnetic phase. 
We find that the composite correlation, which is a measure of differentiation in the
Kondo coupling between the two channels, is enhanced in this dome-shaped region. 
We also find that the specific heat coefficient is enhanced in this region in the paramagnetic side, 
indicating heavy fermion behavior not only in the vicinity of the critical field where the antiferroic-spin 
order vanishes but also in a certain region of the field and temperature. 
We discuss the results in comparison with the ordinary Kondo lattice model. 
We also discuss the implication of our findings to the peculiar behavior observed 
in the 1-2-20 compounds such as $\rm{PrIr_2Zn_{20}}$, $\rm{PrRh_2Zn_{20}}$, and 
$\rm{PrV_2Al_{20}}$ under a magnetic field. 
\end{abstract}

\pacs{Valid PACS appear here}

\maketitle

\section{Introduction}
\label{introduction}

$f$-electron systems provide 
a good playground for competition between itinerant and localized nature of electrons~\cite{RevModPhys.56.755,RevModPhys.73.797,hewson_1993}. 
In these systems, $f$ electrons comprise localized degrees of freedom whose nature depends on the ground-state multiplet determined by electron correlations, the spin-orbit coupling, and the crystalline electric field (CEF). 
When the multiplet is a Kramers doublet, the localized degrees of freedom are described by pseudo-spins. 
In this case, the hybridization between the localized magnetic moments and 
conduction electrons yields two competing interactions: 
the Ruderman-Kittel-Kasuya-Yosida (RKKY) interaction~\cite{PhysRev.96.99,10.1143/PTP.16.45,PhysRev.106.893} 
and the Kondo coupling~\cite{10.1143/PTP.32.37}. 
The RKKY interaction is an effective exchange interaction between the localized moments mediated by the conduction electrons, 
which favors magnetic ordering. 
Meanwhile, the Kondo coupling is the bare coupling between the  
conduction electron spins and the localized magnetic moments, 
which prefers singlet formation between the 
conduction electrons and the localized moments. 
The competition between these two leads to the so-called Doniach phase diagram, where 
a magnetically ordered phase meets with a paramagnetic (PM) phase at a quantum critical point (QCP)~\cite{DONIACH1977231}. 
In the PM region, the singlet formation by the Kondo coupling leads to a logarithmic temperature dependence of the electrical resistivity and heavy fermion (HF) behavior with strong enhancement of the effective electron mass at low temperature, 
which is called the Kondo effect~\cite{RevModPhys.56.755}. 
In addition, unusual behaviors, such as non-Fermi liquid (NFL) behavior and superconductivity, 
are observed near the QCP~\cite{RevModPhys.79.1015,Gegenwart2008,Si1161,HEUSER1999392}. 
This picture called the Doniach paradigm provides comprehensive understanding of many $f$-electron systems with Kramers doublet, 
while several interesting behaviors beyond it have been discussed, e.g., in the presence of magnetic frustration~\cite{SI200623,Coleman2010}.  

On the other hand, when the ground-state multiplet of the localized degrees of freedom is a non-Kramers doublet, the system  
has orbital degrees of freedom described by multipole operators, instead of the magnetic degrees of freedom. 
In particular, when the multiplet is a $\Gamma_3$ doublet, 
which may appear in, e.g., U$^{4+}$ and Pr$^{3+}$ ions under the cubic CEF, 
the system has quadrupole degree of freedom. 
The possibility of the Kondo effect caused by the coupling to the quadrupole degree of freedom was first proposed by Cox, 
by introducing a two channel model~\cite{PhysRevLett.59.1240}. 
In the subsequent studies, the two-channel Kondo model with a single quadrupole impurity was solved, e.g., by 
the Bethe ansatz~\cite{SACRAMENTO1989245} and the conformal field theory~\cite{PhysRevB.48.7297}, 
which revealed that overscreening of the quadrupole by 
conduction electrons leads to NFL behavior. 
It leads to anomalous temperature ($T$) dependence at low
temperature, such as $\ln T$ in the specific heat and the magnetic susceptibility, 
and $\sqrt{T}$ in the electrical resistivity~\cite{PhysRevLett.67.3160, PhysRevB.48.7297}. 

Besides the impurity problem with a single quadrupole, the two-channel Kondo lattice (TCKL) model, 
which has the quadrupole degree of freedom at each lattice site, 
has also been studied for understanding of the systems with dense quadrupoles such as UBe$_{13}$. 
For instance, numerical studies based on the dynamical mean-field theory (DMFT)~\cite{RevModPhys.68.13} 
revealed NFL behavior at low 
temperature when the system remains as a PM state~\cite{PhysRevLett.77.1612}. 
The possibility of symmetry breaking was also pointed out, for instance, 
an antiferromagnetic state, an odd-frequency superconducting state~\cite{PhysRevLett.78.1996}, 
and a channel-symmetry broken state~\cite{0953-8984-20-25-255231}. 
(Note that the quadrupole and spin degrees of freedom are described by pseudo-spins and channels, respectively, following the convention by Cox~\cite{PhysRevLett.59.1240}.)
More recently, the DMFT combined with the continuous-time quantum Monte Carlo (CTQMC) method~\cite{RevModPhys.83.349,doi:10.1143/JPSJ.76.114707} 
uncovered the phase diagram of the TCKL model while changing electron filling and temperature.
Among many phases, an interesting state appears with a ferroic
channel order in doped regions close to half filling;  
it is regarded as a ``composite state'' in which electrons in one of the two channels form ``spin-singlet'' with local moments 
while electrons in the other channel 
are rather free~\cite{PhysRevLett.107.247202,doi:10.7566/JPSJ.82.044707,PhysRevLett.112.167204}.
Thus, the phase diagram for the quadrupole Kondo systems 
is richer than the Doniach phase diagram for the ordinary Kondo system due to the 
interplay between quadrupole and spin degrees of freedom. 

Recent discovery of a family of compounds called Pr-based 1-2-20 systems 
has brought further progress in the research of the quadrupole Kondo systems~\cite{PhysRevLett.106.177001,PhysRevB.87.205106,doi:10.1143/JPSJ.80.063701,doi:10.7566/JPSJ.85.082002}.
The chemical formula is given by Pr$Tr_2X_{20}$, where $Tr$ is a transition metal ion and $X$ is Zn or Al. 
In these materials, the Pr$^{3+}$ cation is 
surrounded by the 16 Zn or Al ions, which leads to 
the $\Gamma_3$ non-Kramers doublet under the cubic CEF and 
strong coupling between the quadrupoles and conduction electrons. 
Indeed, an antiferroic-quadrupole (AFQ) order was observed at low 
temperature and NFL behavior was found above the critical temperature 
in PrIr$_2$Zn$_{20}$~\cite{PhysRevLett.106.177001}, PrRh$_2$Zn$_{20}$~\cite{PhysRevB.87.205106}, 
and PrV$_2$Al$_{20}$~\cite{doi:10.1143/JPSJ.80.063701}. 
The NFL behavior is in good agreement with the theoretical results obtained by the $1/N$ expansion for 
the two-channel Anderson lattice model~\cite{doi:10.7566/JPSJ.84.114714}. 
In addition, in Y$_{1-x}$Pr$_x$Ir$_2$Zn$_{20}$ where Pr is diluted by nonmagnetic Y, 
the temperature dependences of the electrical resistivity and the specific heat are well explained by the impurity 
two-channel Kondo model~\cite{PhysRevLett.121.077206}. 
In the diluted materials, quadrupole fluctuations were also observed in ultrasonic measurements~\cite{PhysRevLett.123.067201}. 
Although the magnetic Kondo effect may compete with the quadrupole Kondo effect in these non-Kramers systems, 
a recent theoretical study pointed out that the quadrupole interactions play 
a dominant role in $\rm PrIr_2Zn_ {20}$ and $\rm PrRh_2Zn_{20}$~\cite{doi:10.7566/JPSJ.85.064708}. 

Interestingly, the 1-2-20 compounds exhibit peculiar behaviors under a magnetic field. 
For instance, $\rm PrIr_2Zn_{20}$ and $\rm PrRh_2Zn_{20}$ show HF behavior 
in a certain range of the magnetic field 
where the AFQ order is suppressed~\cite{PhysRevB.94.075134,doi:10.7566/JPSJ.86.044711}. 
Anomalous enhancement of the Seebeck coefficient was also reported~\cite{doi:10.7566/JPSCP.3.011091,doi:10.7566/JPSJ.86.044711}. 
Furthermore, the thermal expansion for PrIr$_2$Zn$_{20}$ in this field region cannot be explained by a CEF model, 
whereas the higher-field behavior is well accounted for~\cite{PhysRevB.99.081117}. 
On the other hand, $\rm PrV_2Al_{20}$ shows anomalous enhancement of the resistivity around the 
critical field where the quadrupole order disappears~\cite{PhysRevB.91.241102}. 
The temperature dependence deviates from the scaling relation of NFL behavior expected from the two-channel Anderson model 
below $8$~K in a certain range of the magnetic field~\cite{doi:10.7566/JPSJ.89.013704}. 
These experimental results suggest that the 1-2-20 systems exhibit unconventional behaviors not only near the 
critical field but also in a certain range of temperature and magnetic fields. 
This is in stark contrast to the ordinary Kramers Kondo systems 
where the NFL behavior is limited to a narrow critical region in the vicinity of the QCP. 

Theoretical understanding is, however, still limited for the field effects on the non-Kramers quadrupole systems. 
The magnetic phase diagrams were studied by a mean-field calculation~\cite{doi:10.7566/JPSJ.83.034709}, 
classical Monte Carlo simulations~\cite{doi:10.7566/JPSJ.85.094001,PhysRevB.97.115111}, 
and the Landau theory~\cite{PhysRevB.98.134447}. 
These analyses were performed for effective models describing the quadrupole degree of freedom. 
Models explicitly including the coupling to conduction electrons were also studied 
by slave-particle mean-field approximations and the Landau theory~\cite{PhysRevB.98.235143,PhysRevB.100.205122,PhysRevB.101.075133}. 
However, effects of quantum fluctuations arising from the interplay between quadrupole and spin degrees of freedom as well as spatial fluctuations 
have not been fully elucidated thus far, 
despite their importance for understanding of not only the magnetic phase diagram but 
also the unconventional NFL and HF behaviors observed in experiments. 

In this paper, we study the TCKL model in a magnetic field by using a cluster extension of the DMFT (CDMFT) 
combined with the CTQMC method as the impurity solver. 
The cluster extension enables us to 
study the effect of magnetic fields on the competition and cooperation between the spin and quadrupole degrees of freedom,
taking into account spatial correlations beyond the previous studies by the single-site DMFT
\cite{PhysRevLett.78.1996,PhysRevLett.107.247202,doi:10.7566/JPSJ.82.044707,PhysRevLett.112.167204}. 
Our model includes two types of the magnetic fields: One represents the ordinary Zeeman coupling for the conduction electrons, 
and the other an effective coupling to the quadrupole degree of freedom 
through the modulation of the CEF splitting. 
Performing extensive numerical simulations for the two types of the fields, 
we show that the antiferroic-spin (AF-spin) ordered phase 
(corresponding to the AFQ ordered phase in non-Kramers quadrupole systems) is suppressed while increasing the magnetic fields, 
and eventually, the system shows FL behavior in the high-field PM state. 
In the intermediate region, we find an interesting 
state in which one of the two channels shows NFL behavior while the other remains as FL. 
We call this the channel-selective NFL (CS-NFL) state. 
Interestingly, the CS-NFL state appears in a dome-shaped region 
that extends 
from the AF-spin ordered state to the PM state.
We find that the composite correlation is enhanced in this region, and 
in addition, the specific heat coefficient is also enhanced on the paramagnetic side. 
Thus, the HF behavior is found not only near the critical field where the
AF-spin order disappears but also in a certain range of field and temperature. 
We discuss our findings in comparison with the ordinary Kramers Kondo system and the unconventional behaviors discovered in the 1-2-20 systems.

This paper is organized as follows.
In Sec.~\ref{sec:modelandmethod}, we introduce the model and method used in this study. 
We also introduce the definitions of the physical quantities and how to compute them. 
We present the results in Sec.~\ref{sec:results}. 
First, we display the phase diagram determined by the
AF-spin order parameter and the specific heat in Sec.~\ref{subsec:phasediagram}. 
Next, we discuss the composite 
correlation and the relation to the enhancement of the specific heat coefficient in Sec.~\ref{subsec:compositefluctiationandspecificheatcoefficient}. 
Then, in Sec.~\ref{subsec:channelselectivenonfermiliquidbehavior}, 
we unveil the CS-NFL behavior by analyzing the temperature dependence of the self-energy. 
We also study how these results depend on the Kondo coupling in Sec.~\ref{subsec:jdependences}.
In Sec.~\ref{sec:discussion}, we discuss the results for the TCKL model, 
in comparison with those for the ordinary Kondo lattice model (Sec.~\ref{subsec:magneticphasediagramoftcklmodel}) 
and also with the experimental results for the 1-2-20 systems (Sec.~\ref{subsec:therelationtothe1220systems}).  
Finally, Sec.~\ref{sec:summary} is devoted to the summary.

\section{Model and Method}
\label{sec:modelandmethod}

\subsection{Model}
\label{subsec:model}

We study the TCKL model under a magnetic field, whose Hamiltonian is given by
\begin{eqnarray}
\mathcal{H}  & = &  \mathcal{H}_{\rm TCKL} + \mathcal{H}_{\rm Zeeman} +  \mathcal{H}_{\rm CEF}.
\label{eq:H_total}
\end{eqnarray}
The first term describes the electrons coupled to the quadrupole degree of freedom as 
\begin{eqnarray}
\mathcal{H}_{\rm TCKL} & = & \sum_{\bm{k}\alpha \sigma} (\epsilon_{\bm{k}} - \mu) 
c_{\bm{k}\alpha \sigma}^{\dagger}c_{\bm{k}\alpha \sigma}  + J\sum_{i\alpha} \bm{s}_{i\alpha} \cdot \bm{S}_i, 
\label{eq:H_TCKL}
\end{eqnarray}
where $c_{\bm{k}\alpha \sigma}$ ($c_{\bm{k}\alpha \sigma}^{\dagger}$) is an annihilation (creation) operator of a conduction electron 
with momentum $\bm{k}$, channel $\alpha=1, 2$, and pseudospin $\sigma=\uparrow, \downarrow$. 
Note that, following the conventions in the previous studies~\cite{PhysRevLett.107.247202,doi:10.7566/JPSJ.82.044707,PhysRevLett.112.167204}, 
the channel $\alpha$ and pseudospin $\sigma$ 
represent the spin and orbital degrees of freedom in real systems, respectively. 
The first term in Eq.~(\ref{eq:H_TCKL}) describes the kinetic energy of the conduction electrons. 
For simplicity, we assume the electron hopping only for nearest-neighbor sites on a three-dimensional cubic lattice, 
which yields the dispersion relation 
$\epsilon_{\bm{k}} = -2t \sum_{\gamma=x,y,z} \cos(k_{\gamma}\red{a})$, 
where $\bm{k}=(k_x,k_y,k_z)$ and $a$ is the lattice constant; $\mu$ is the chemical potential. 
We set $6t=1$ as the energy unit and $a=1$ as the length unit. 
The second term in Eq.~(\ref{eq:H_TCKL}) denotes the coupling between the conduction electrons and 
the local quadrupole moments; 
$\bm{s}_{i\alpha} 
= \frac12 \Sigma_{\sigma\sigma'} c_{i\alpha\sigma}^\dagger\boldsymbol{\sigma}_{\sigma\sigma'}c_{i\alpha\sigma'}$ 
is the pseudospin-$\frac{1}{2}$ operator  
for a conduction electron at site $i$ in channel $\alpha$ 
($\boldsymbol{\sigma}$ is the Pauli matrix, and 
$c_{i\alpha\sigma}$ and $c_{i\alpha\sigma}^\dagger$ are Fourier components of 
$c_{\bm{k}\alpha \sigma}$ and $c_{\bm{k}\alpha \sigma}^{\dagger}$, respectively), and 
$\bm{S}_i$ represents another 
pseudospin-$\frac12$ operator representing the local quadrupole degree  
of freedom at site $i$, which we call the local moment hereafter. 
For simplicity, we assume that the interaction is onsite and isotropic in quadrupole space. 
We take $J=0.8$, except in Sec.~\ref{subsec:jdependences}. 

The second and third terms in Eq.~(\ref{eq:H_total}) represent the coupling to 
two types of external magnetic fields.
The second term represents the Zeeman coupling for the conduction electrons, which is given by 
\begin{eqnarray}
\mathcal{H}_{\rm  Zeeman} & = & - \bm{h}_{\rm Zeeman} \cdot \sum_{i\sigma} \tilde{\bm{s}}_{i\sigma},
\end{eqnarray}
where $\tilde{\bm{s}}_{i\sigma} = \frac12 \Sigma_{\alpha\alpha'} c_{i\alpha\sigma}^\dagger\boldsymbol{\sigma}_{\alpha\alpha'}c_{i\alpha\sigma'}$ 
is the spin-$\frac12$ operator of a conduction electron at site $i$ for orbital $\sigma$. 
In the following study, we apply the magnetic field to the $z$ direction, namely, $\bm{h}_{\rm Zeeman} = (0,0,h)$, and hence,
\begin{equation}
\mathcal{H}_{\rm Zeeman} = - h \sum_{i\sigma} \tilde{s}_{i\sigma}^z 
= - \frac{h}{2} \sum_i (n_{i1} - n_{i2}),
\label{eq:H_Zeeman}
\end{equation}
where $\tilde{s}_{i\sigma}^z$ is the $z$ component of the spin and $n_{i\alpha} = \sum_{\sigma} c_{i\alpha\sigma}^\dagger c_{i\alpha\sigma}$ is the number operator 
for the conduction electrons 
in channel $\alpha$ at site $i$. 
Note that Eq.~(\ref{eq:H_Zeeman}) splits the energy of the two channels as the channel degree of freedom in the present model describes the spin in real systems. 

On the other hand, the third term in Eq.~(\ref{eq:H_total}) represents another effect of the magnetic field through the CEF level splitting. 
While the magnetic field does not directly couple to the non-Kramers doublet in the 
ground state, it couples to the magnetic excited states. 
This coupling perturbs the non-Kramers doublet through the CEF, which is effectively described by the Zeeman-like term in the second-order perturbation as
\begin{eqnarray}
\mathcal{H}_{\rm 
\rm CEF} & = & - h_{\rm CEF} \sum_{i} S_i^z.  
\label{eq:H_CEF}
\end{eqnarray}
In the limit of a weak magnetic field $h$, 
$h_{{\rm CEF}}$ is expected to be proportional to $h^2$~\cite{doi:10.7566/JPSJ.83.034709}. 

The two terms, $\mathcal{H}_{\rm Zeeman}$ and $\mathcal{H}_{\rm CEF}$, affect the system in a different way. 
Near half filling (two conduction electrons per site on average), the system prefers 
AF-spin ordering at low temperature in the absence of the magnetic field~\cite{PhysRevLett.112.167204}. 
Although both $\mathcal{H}_{\rm Zeeman}$ and $\mathcal{H}_{\rm CEF}$ destabilize the AF-spin order, 
the former induces a composite order, while the latter simply leads to a PM state;
we show the phase diagram while changing $h$ and $h_{\rm CEF}$ in Appendix~\ref{app:hh}. 
In the following sections, we perform the calculations by assuming 
\begin{eqnarray}
h_{\rm CEF} & = & \frac74h^2,
\label{eq:h}
\end{eqnarray}
for which the system undergoes a transition from the AF-spin ordered state to the PM state without 
going through the composite ordered state stabilized by $\mathcal{H}_{\rm Zeeman}$, as 
shown by the dotted curve in Fig.~\ref{fig:hhb100} in Appendix~\ref{app:hh}.

\subsection{Method}
\label{subsec:method}

We use the CDMFT~\cite{RevModPhys.68.13,PhysRevLett.87.186401} to study the model in Eq.~(\ref{eq:H_total}) 
by taking a two-site cluster composed of neighboring sites on the cubic lattice.
The cluster extension enables us to directly incorporate staggered orders like the AF-spin order under a uniform magnetic field.
We adopt the CTQMC technique as the impurity solver in the CDMFT calculations~\cite{RevModPhys.83.349,doi:10.1143/JPSJ.76.114707}. 
In each CDMFT loop, we perform $2 \times 10^9$ samplings in the CTQMC calculations. 
We obtain the local Green functions by taking the summation over $N= 16^3$ points in momentum space. 
All the calculations are performed by fixing the electron filling 
at $n=\frac{1}{2} \sum_{i\alpha\sigma} \langle c_{i\alpha\sigma}^\dagger c_{i\alpha\sigma}\rangle = 0.9$, 
where the sum of $i$ is taken for the two sites within the cluster. 
This corresponds to $10$\% hole doping from the half filling. 

The self-consistent solution is obtained when the values of the local Green functions 
at each Matsubara frequency converge within the statistical errors. 
For the calculation of the internal energy $\langle \mathcal{H} \rangle$ for the specific heat
[see Eq.~(\ref{eq:c}) below], 
we follow the method described in Appendix of Ref.~\cite{doi:10.7566/JPSJ.82.044707}, 
by performing additional 10 CDMFT loops with $ 5\times 10^9$ samplings and another 10 loops with $1\times 10^{10}$ 
samplings for sufficient precision.

\subsection{Physical observables}
\label{subsec:thephysicalquantities}

We discuss the finite-temperature properties of the model in Eq.~(\ref{eq:H_total}) 
by calculating the following physical quantities. 
First, to identify the AF-spin ordered phase, we introduce 
the AF-spin order parameter for the conduction electrons, which is defined by
\begin{eqnarray}
m_{\rm{AF}} &=&   \sum_{\alpha} \left|\langle \bm{s}_{1\alpha}-\bm{s}_{2\alpha}\rangle\right|, 
\label{eq:af} 
\end{eqnarray}
where $\bm{s}_{1\alpha}$ and $\bm{s}_{2\alpha}$ represent the pseudospins at site 1 and 2 
within the two-site cluster, respectively.  
We also compute the specific heat per site by taking $T$ derivative of the internal energy as 
\begin{eqnarray}
C =  \frac1N \frac{d\braket{\mathcal{H}}}{dT}, 
\label{eq:c}
\end{eqnarray}
where $\braket{\mathcal{H}}$ is calculated by the method described in Appendix of Ref.~\cite{doi:10.7566/JPSJ.82.044707}. 

In addition, following Ref.~\cite{PhysRevLett.107.247202} we calculate the composite correlation, 
which is defined by
\begin{eqnarray}
\Psi &=&  \frac{1}{2}\sum_{i=1}^2 \langle  (\bm{s}_{i1} - \bm{s}_{i2}) \cdot \bm{S}_i
\rangle. 
\label{eq:psi}
\end{eqnarray}
This quantity measures differentiation in the couplings of the two channels to the local moment. 
As $\mathcal{H}_{\rm Zeeman}$ is a symmetry breaking field for the the channel degree of freedom, 
$\Psi$ becomes always nonzero under the magnetic field. 
Nonetheless, it plays an important role in this study to discuss the CS-NFL behavior. 

Furthermore, in order to distinguish the FL and NFL behaviors, 
we analyze the imaginary part of the self-energy. 
In the CDMFT, the self-energy on the lattice is 
related with that within the cluster as~\cite{PhysRevB.85.035102} 
\begin{eqnarray}
\Sigma^{
\lambda\lambda^\prime}(\bm{k},i\omega_n) = 
\frac{1}{2}\sum^{2}_{i,j=1} \Sigma^{\lambda\lambda^\prime}_{ij} 
(i\omega_n){\rm e}^{i\bm{k}\cdot(\bm{r}_i - \bm{r}_j)}, 
\label{eq:sigma_l}
\end{eqnarray} 
where $\lambda=(\alpha\sigma)$ 
and $\omega_n = \pi T(2n+1)$ is Matsubara frequency; $\bm{r}_i$ 
denotes the position vector for site $i$ within the cluster.
We calculate Eq.~(\ref{eq:sigma_l}) only for $\bm{k}=\bm{0}$, 
while we confirm the results for $\bm{k}=(\pi,\pi,\pi)$ are qualitatively the same. 
In Sec.~\ref{subsec:channelselectivenonfermiliquidbehavior}, 
we show the results for the diagonal sums in each channel as 
\begin{equation}
\Sigma_{\alpha}(i\omega_n) = \sum_\sigma \Sigma^{\lambda\lambda}(\bm{0}, i\omega_n).
\end{equation}
The FL theory predicts that the retarded self-energy satisfies 
${\rm Im}\Sigma^R (\omega,T) \propto \omega^2 + \pi^2 T^2$ and ${\rm Re}\Sigma^R(\omega,T) \propto \omega$ 
for $\omega \to 0$ in three dimensions. 
Hence, the self-energy at the smallest Matsubara frequency $\omega_0 = \pi T$ is expected to 
behaves as ${\rm Re}\Sigma (i\omega_0) \propto T^3$ 
and ${\rm Im}\Sigma (i\omega_0) \propto T$ at low temperature~\cite{PhysRevB.86.155136}. 
To measure the deviation from this FL behavior, 
we estimate the power $\nu_{\alpha}$ defined as 
\begin{eqnarray}
{\rm Im}\Sigma_{\alpha}(i\omega_0) \propto T^{\nu_{\alpha}}
\label{eq:nu}
\end{eqnarray}
by fitting the $T$ dependence of ${\rm Im}\Sigma_\alpha(i\omega_0)$. 
The value of $\nu_\alpha$ characterizes the nature of the system:  
FL for $\nu_\alpha \sim 1$, 
NFL for $0<\nu_\alpha<1$, and an insulator for $\nu_\alpha<0$.

\section{Results}
\label{sec:results}

In this section, we present our CDMFT results. 
We show the $h$-$T$ phase diagram 
in Sec.~\ref{subsec:phasediagram} and enhancement of the composite 
correlation and the specific heat in Sec.~\ref{subsec:compositefluctiationandspecificheatcoefficient}. 
We elaborate on the CS-NFL behavior 
by analyzing the imaginary part of the self-energy in Sec.~\ref{subsec:channelselectivenonfermiliquidbehavior}. 
We also discuss the $J$ dependence of the phase diagram in Sec.~\ref{subsec:jdependences}. 

\subsection{Phase diagram}
\label{subsec:phasediagram}

\begin{figure}[t]
\begin{tabular}{cc}
      \begin{minipage}[t]{1.0\hsize}
        \centering
        \includegraphics[width=8.5cm,pagebox=cropbox,clip]{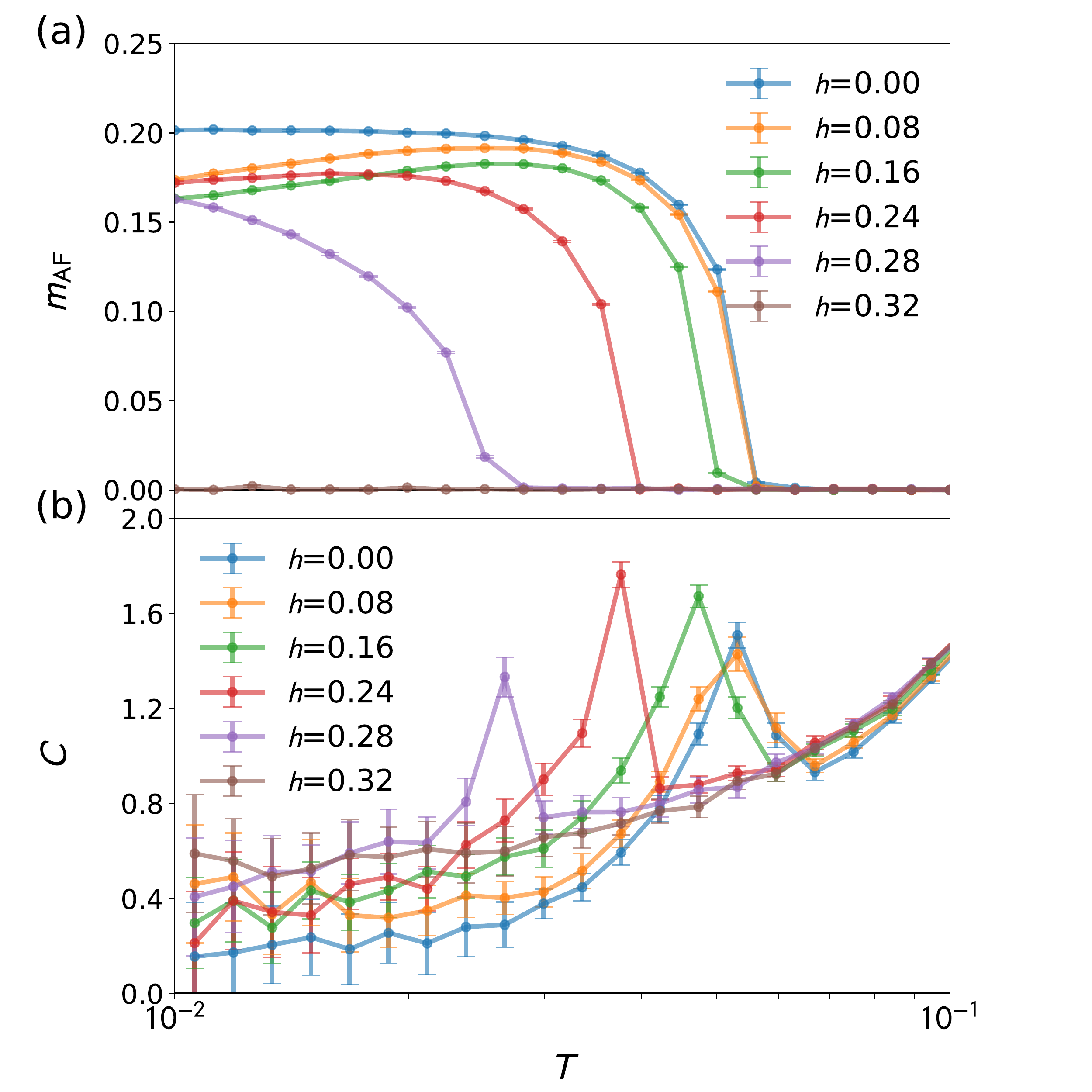}
      \end{minipage} \\
      \begin{minipage}[t]{1.0\hsize}
        \centering
        \includegraphics[width=8.5cm,pagebox=cropbox,clip]{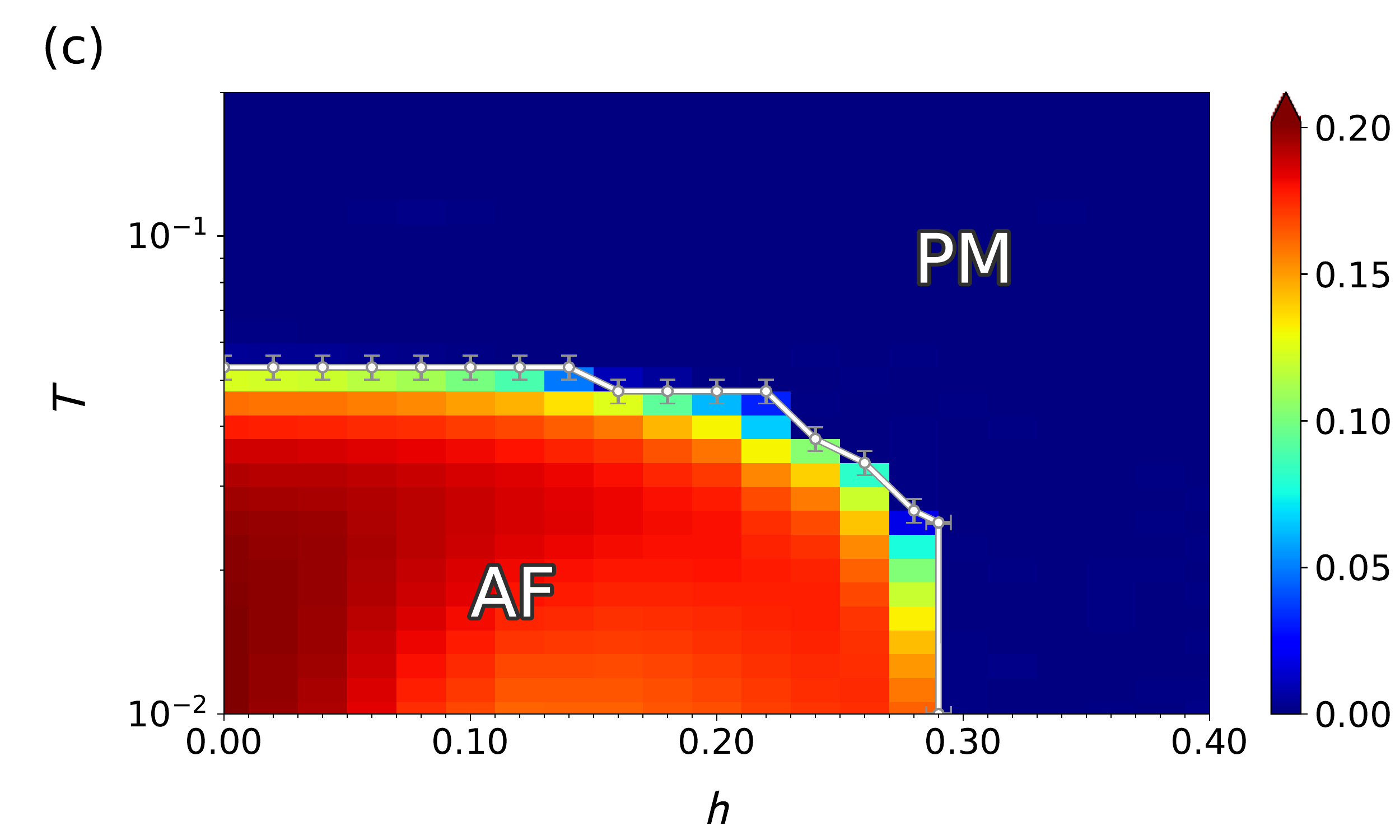}
        \caption{
Temperature dependences of (a) the AF-spin order parameter 
$m_{\rm AF}$ in Eq.~(\ref{eq:af}) and 
(b) the specific heat per site 
$C$  in Eq.~(\ref{eq:c}) for several values of $h$. 
(c) Phase diagram of the model in Eq.~(\ref{eq:H_total}) as a function of $h$ and $T$.
The contour plot shows the value of $m_{\rm  AF}$.
The white line indicates the phase boundary for the
AF-spin ordered phase connecting the onset temperature of $m_{\rm AF}$. 
The data are calculated for $J=0.8$ and $n=0.9$. 
                }
        \label{Fig:afc}
      \end{minipage}
\end{tabular}
\end{figure}

Figure~\ref{Fig:afc}(a) shows 
the $T$ dependence of the AF-spin order parameter $m_{\rm AF}$ 
defined in Eq.~(\ref{eq:af}) for several values of the magnetic field $h$. 
At zero field, $m_{\rm AF}$ becomes nonzero below $T\simeq 0.055$. 
While increasing $h$, the onset temperature decreases and vanishes to zero at $h\simeq 0.30$. 
Correspondingly, the specific heat per site $C$ defined in Eq.~(\ref{eq:c}) 
exhibits a sharp peak at the same temperature, 
as shown in Fig.~\ref{Fig:afc}(b). 
The results indicate that the system exhibits a phase transition 
from the high-temperature PM state to the low-temperature
AF-spin ordered state in the region of $h \lesssim 0.30$. 

The phase diagram is shown as a function of $h$ and $T$ 
in Fig.~\ref{Fig:afc}(c). 
The contour color represents the value of $m_{\rm AF}$ and the 
gray line indicates the phase boundary 
connecting the onset temperature of $m_{\rm AF}$. 
The result shows that the AF-spin ordered phase is realized 
in the low-$T$ and low-$h$ region.
In the region for $h\lesssim 0.28$, $m_{\rm AF}$ appears to grow continuously while lowering $T$, 
suggesting a continuous phase transition. 
On the other hand, on the verge at $h\simeq 0.30$ where the phase boundary is almost vertical, 
the transition might be turned into a discontinuous one. 
We will return to this point in Sec.~\ref{subsec:compositefluctiationandspecificheatcoefficient}. 

We note that $m_{\rm AF}$ slightly decreases with $T$ in the
AF-spin ordered phase under the magnetic field; see the plots for $h=0.08$, $0.16$, and $h=0.24$ in Fig.~\ref{Fig:afc}(a). 
This is due to singlet formation in one of the two channels under the magnetic field as follows. 
While increasing $h$, the electron density in the channel $1$ is increased due to the Zeeman coupling in Eq.~(\ref{eq:H_Zeeman}) 
and becomes close to half filling $\langle n_{i1} \rangle \simeq 1$. 
Then, the coupling $J$ prefers singlet formation between 
$\bm{s}_{i1}$ and $\bm{S}_i$, which reduces the local moment of the conduction electrons, 
$\bm{s}_{i1}$, and consequently, $m_{\rm AF}$. 
This is indeed confirmed by the enhancement of $\Psi$ discussed in the next section. 

\subsection{Composite correlation and heavy fermion behavior}
\label{subsec:compositefluctiationandspecificheatcoefficient}

\begin{figure}[htbp]
\begin{flushleft}
\begin{tabular}{cc}
      \begin{minipage}[t]{1.0\hsize}
        \centering
        \includegraphics[width=8.5cm,pagebox=cropbox,clip]{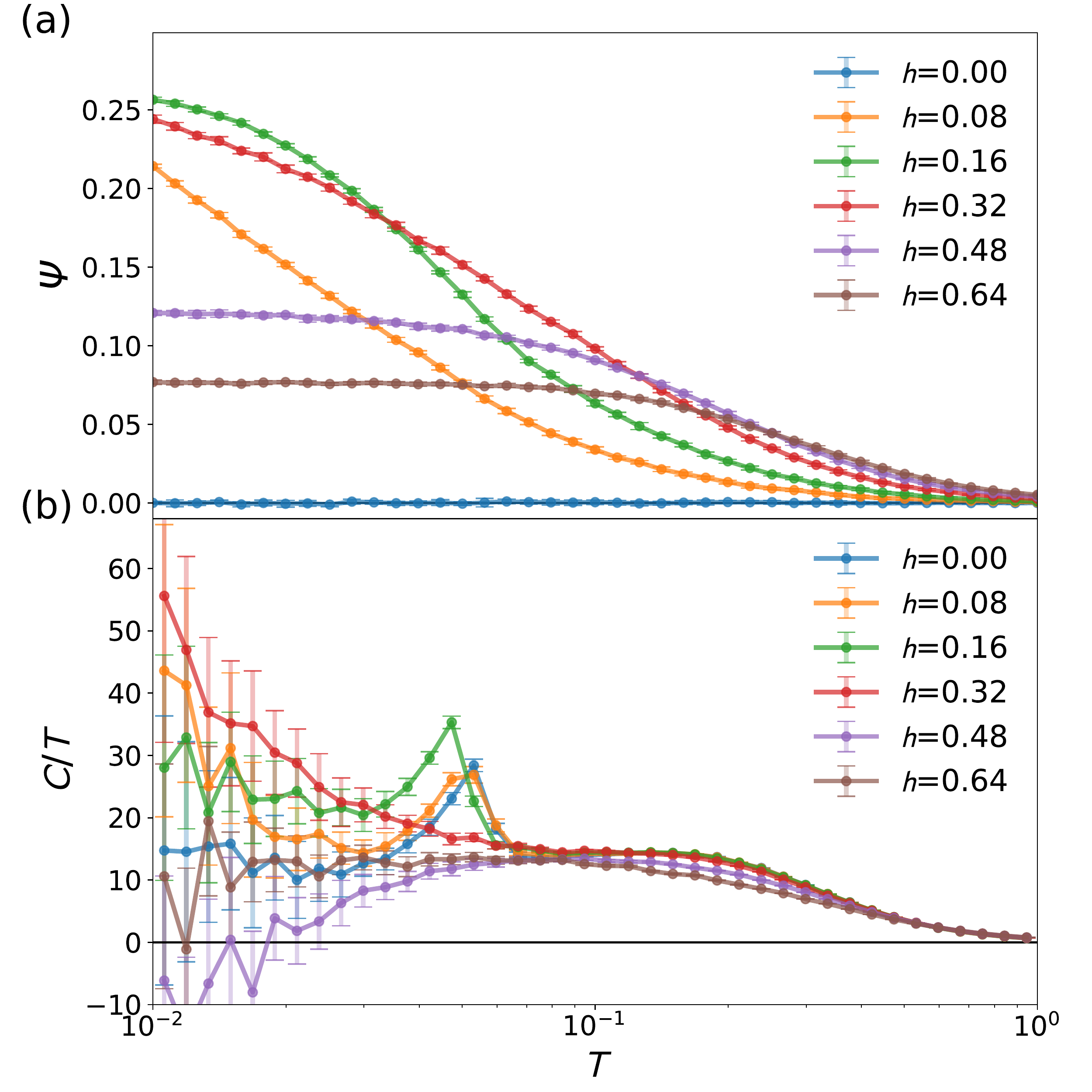}
        \caption{
Temperature dependences of (a) 
$\Psi$ in Eq.~(\ref{eq:psi}) and (b) 
$C$ in Eq.~(\ref{eq:c}) divided by $T$ for several $h$. 
The data are calculated for $J=0.8$ and $n=0.9$. 
        }
        \label{Fig:ct_phi_t}
      \end{minipage}
\end{tabular}
\end{flushleft}
\end{figure}

Next, we show the results of the composite correlation $\Psi$ 
[Eq.~(\ref{eq:psi})] in Fig.~\ref{Fig:ct_phi_t}(a). 
Since $\Psi$ measures the difference between the couplings 
to the local moment in the two channels, it is zero at $h=0$ where the two channels are equivalent. 
For $h>0$, however, $\Psi$ becomes nonzero even in the high-temperature PM state, as the Zeeman term in Eq.~(\ref{eq:H_Zeeman}) 
is a symmetry breaking field for the channel degree of freedom. 
As shown in Fig.~\ref{Fig:ct_phi_t}(a), $\Psi$ increases as decreasing $T$. 
The low-$T$ values of $\Psi$ is largely enhanced in the intermediate-$h$ region, and reduced for larger $h$. 
This enhancement is related with the decrease of $m_{\rm  AF}$ mentioned in the end of the previous section: 
$\Psi$ becomes large when the channel $1$ 
approaches half filling and the singlet formation is promoted. 
This in turn reduces the coupling to the channel $2$ and leaves 
the conduction electrons in channel $2$ more freely down to low temperature. 

In the ordinary Kondo systems, the coupling to local moments leads to HF behavior at low temperature. 
To examine such behavior, we plot $C$ divided by $T$ in Fig.~\ref{Fig:ct_phi_t}(b). 
Note that the integral of $C/T$ in terms of $T$ gives the entropy per site, and $C/T$ is called the specific heat coefficient 
giving a measure of the effective electron mass in the low-temperature limit. 
The result indicates that $C/T$ is enhanced in the field region where $\Psi$ becomes large. 
This means that the system retains larger residual entropy at low 
temperature in this intermediate-$h$ region. 

\begin{figure}[t]
\begin{flushleft}
\begin{tabular}{cc}
      \begin{minipage}[t]{1.0\hsize}
        \centering
        \includegraphics[width=8.5cm,pagebox=cropbox,clip]{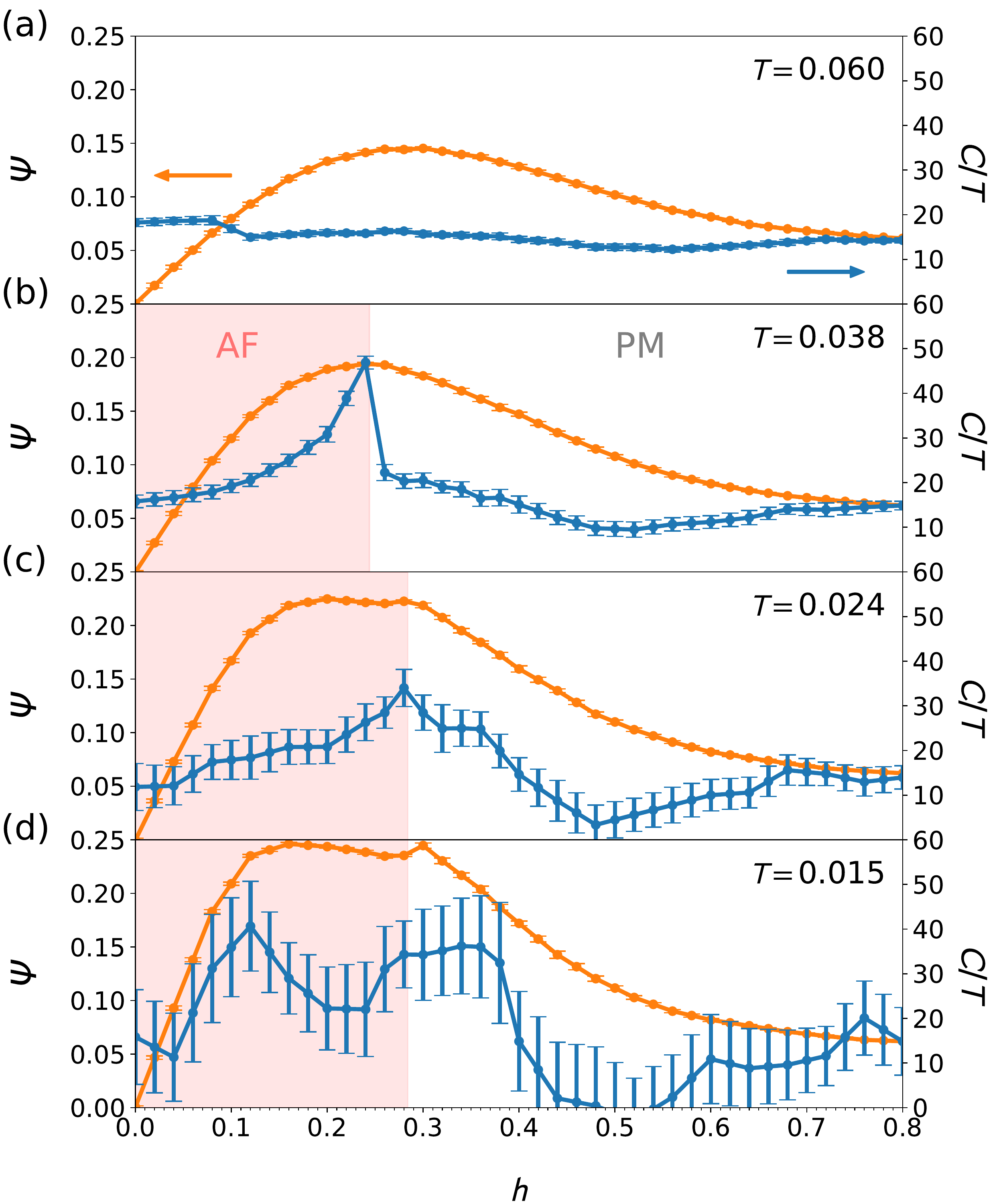}
        \caption{
        $h$ dependences of $\Psi$ and $C/T$ at several $T$. 
        The red shaded regions in (b)-(d) represent the 
AF-spin ordered phase, while the other white regions are the PM phase. 
        }
        \label{Fig:ct_phi_h}
      \end{minipage}
\end{tabular}
\end{flushleft}
\end{figure}

To examine the correlation between $\Psi$ and $C/T$ more carefully,
we plot their $h$ dependences at several $T$ in Fig.~\ref{Fig:ct_phi_h}. 
At all temperatures plotted here, $\Psi$ has a broad peak in the intermediate-$h$ region, 
except for small anomalies at the phase transition from the AF-spin ordered phase to the PM phase 
in Figs.~\ref{Fig:ct_phi_h}(c) and \ref{Fig:ct_phi_h}(d). 
The height of the broad peak grows gradually as decreasing temperature. 
On the other hand, $C/T$ shows more notable temperature dependence. 
At sufficiently high temperature in the PM phase, $C/T$ 
is almost flat as a function of $h$, as exemplified in Fig.~\ref{Fig:ct_phi_h}(a). 
Below $T\simeq 0.055$, $C/T$ shows a sharp peak corresponding to the phase transition, 
as shown in Fig.~\ref{Fig:ct_phi_h}(b) for $T=0.038$. 
At lower temperatures, $C/T$ increases not only in the AF-spin ordered phase but also in the PM phase. 
Thus, the overall $h$ dependence of $C/T$ at low temperature looks similar to that of $\Psi$, 
as discussed in Fig.~\ref{Fig:ct_phi_t}. 
We note, however, that $C/T$ has more complicated $h$ dependence than $\Psi$: 
It shows two humps at the lowest temperature, one inside the AF-spin ordered phase at $h\sim 0.1$ 
and the other in the PM phase at $h\sim 0.3$-$0.4$. 
The former appears in the region where the channel 1 prefers singlet formation, 
while the latter corresponds to the HF behavior discussed later in relation to the CS-NFL state in Sec.~\ref{subsec:channelselectivenonfermiliquidbehavior}. 
We also note that, in the higher-$h$ region, $\Psi$ and $C/T$ exhibit different behavior as shown 
in Figs.~\ref{Fig:ct_phi_h}(c) and \ref{Fig:ct_phi_h}(d);
while $\Psi$ decreases monotonically as increasing $h$, $C/T$ shows a minimum and increases gradually for higher $h$. 
This will also be discussed in Sec.~\ref{subsec:channelselectivenonfermiliquidbehavior}. 

The small anomaly of $\Psi$ at low temperature in Figs.~\ref{Fig:ct_phi_h}(c) and \ref{Fig:ct_phi_h}(d) 
appears to signal a discontinuous change of $\Psi$. 
$m_{\rm AF}$ also shows a jump while changing $h$ in the low-temperature region (not shown). 
This is also consistent with the behavior of $C/T$; while $C/T$ is enhanced 
at relatively high $T$ and $h\lesssim 0.28$ as shown in Fig.~\ref{Fig:ct_phi_h}(b), 
such an anomaly disappears at lower $T$ and higher $h$. 
Thus, all these observations suggest that the AF-spin ordering transition is of second order in the high-$T$ and low-$h$ region, 
but it turns into a first-order one in the low-$T$ region near $h\sim 0.30$, 
while it is difficult to precisely locate the tricritical point between them.

\begin{figure}[htbp]
\begin{flushleft}
\begin{tabular}{cc}
      \begin{minipage}[t]{1.0\hsize}
        \centering
        \includegraphics[width=8.5cm,pagebox=cropbox,clip]{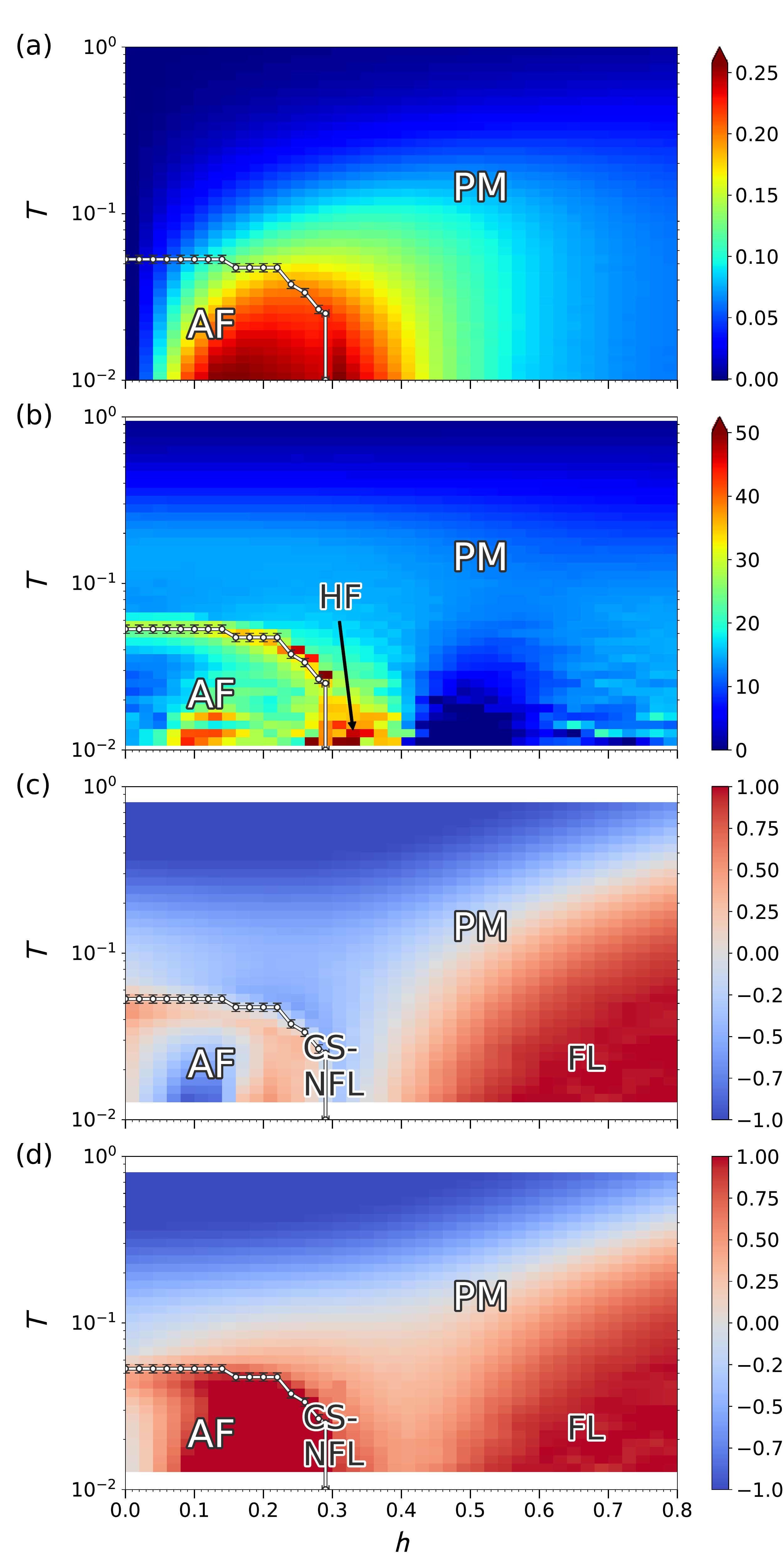}
        \caption{
Contour plots of (a) $\Psi$, (b) $C$/$T$, 
the power of $T$ dependences of Im$ \Sigma_{\alpha}(i\omega_{0})$, $\nu_\alpha$, 
for the channel (c) $\alpha=1$ and (d) $\alpha=2$ [see Eq.~(\ref{eq:nu})]. 
The white lines indicate the phase boundary for the 
AF-spin ordered phase as in Fig.~\ref{Fig:afc}(c). 
}
        \label{Fig:pd4}
      \end{minipage}
\end{tabular}
\end{flushleft}
\end{figure}

We summarize the $h$ and $T$ dependences of $\Psi$ and $C/T$ in Figs.~\ref{Fig:pd4}(a) and \ref{Fig:pd4}(b), respectively. 
As shown in Fig.~\ref{Fig:pd4}(a), $\Psi$ is enhanced in a dome-shaped region. 
The dome extends from inside of the AF-spin ordered phase to the outside PM phase, 
and $\Psi$ changes smoothly across the phase boundary, 
besides the small anomaly associated with the possibly first-order phase transition 
in the low-temperature region. 
On the other hand, $C/T$ is also enhanced in a similar region, but most pronounced in a more limited area for 
$0.30\lesssim h\lesssim 0.40$ below $T\simeq 0.02$ in the PM phase.
We also note that $C/T$ becomes large 
along the second-order phase boundary for smaller $h$ and 
at $h\sim 0.10$ inside the AF-spin ordered phase discussed above. 
We will discuss the pronounced enhancement of $C/T$ in relation to the CS-NFL behavior in the next section.

\subsection{Channel-selective non-Fermi liquid behavior}
\label{subsec:channelselectivenonfermiliquidbehavior}

\begin{figure}[htbp]
\begin{flushleft}
\begin{tabular}{cc}
      \begin{minipage}[t]{1.0\hsize}
        \centering
        \includegraphics[width=8.5cm,pagebox=cropbox,clip]{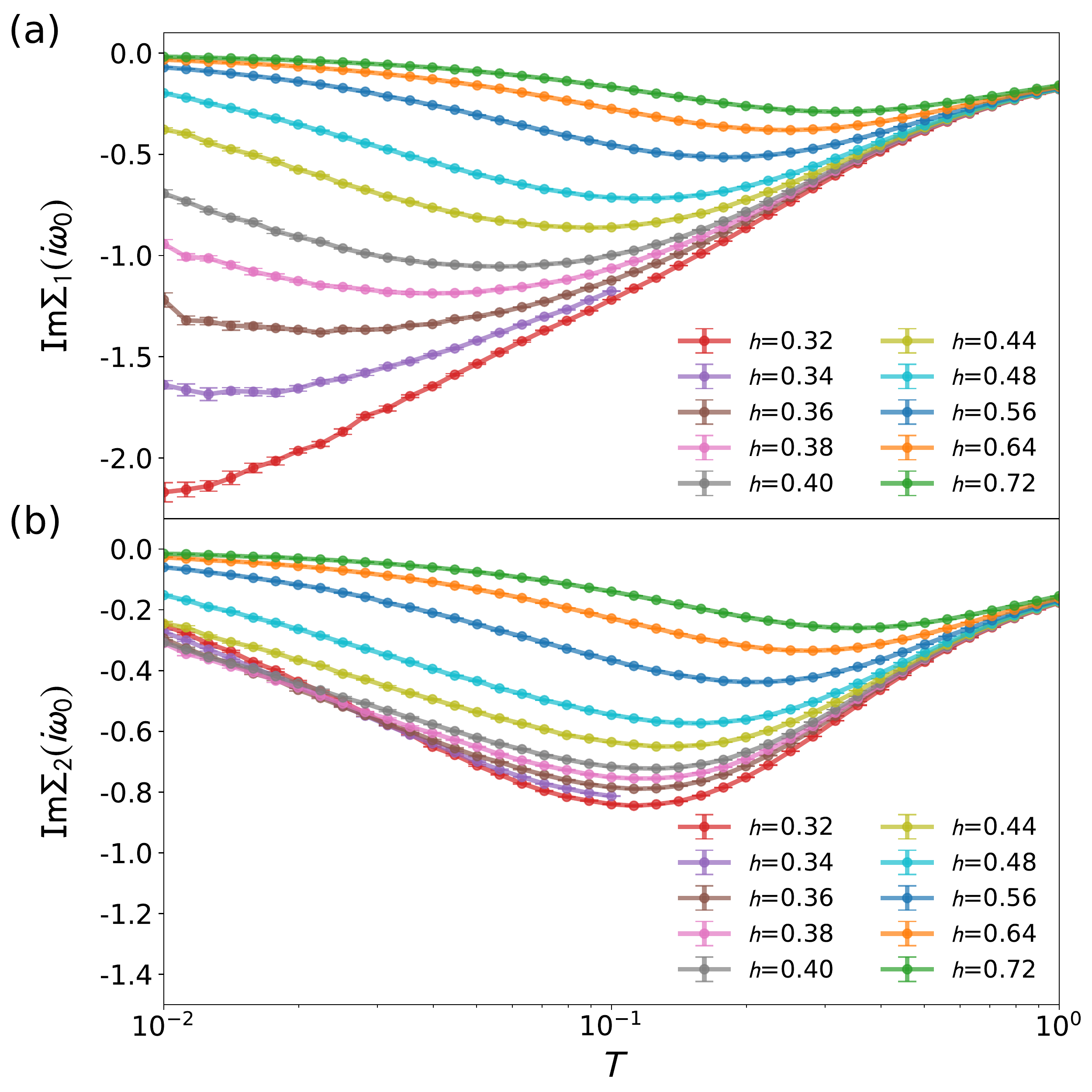}
      \end{minipage} \\
      \begin{minipage}[t]{1.0\hsize}
        \centering
        \includegraphics[width=8.5cm,pagebox=cropbox,clip]{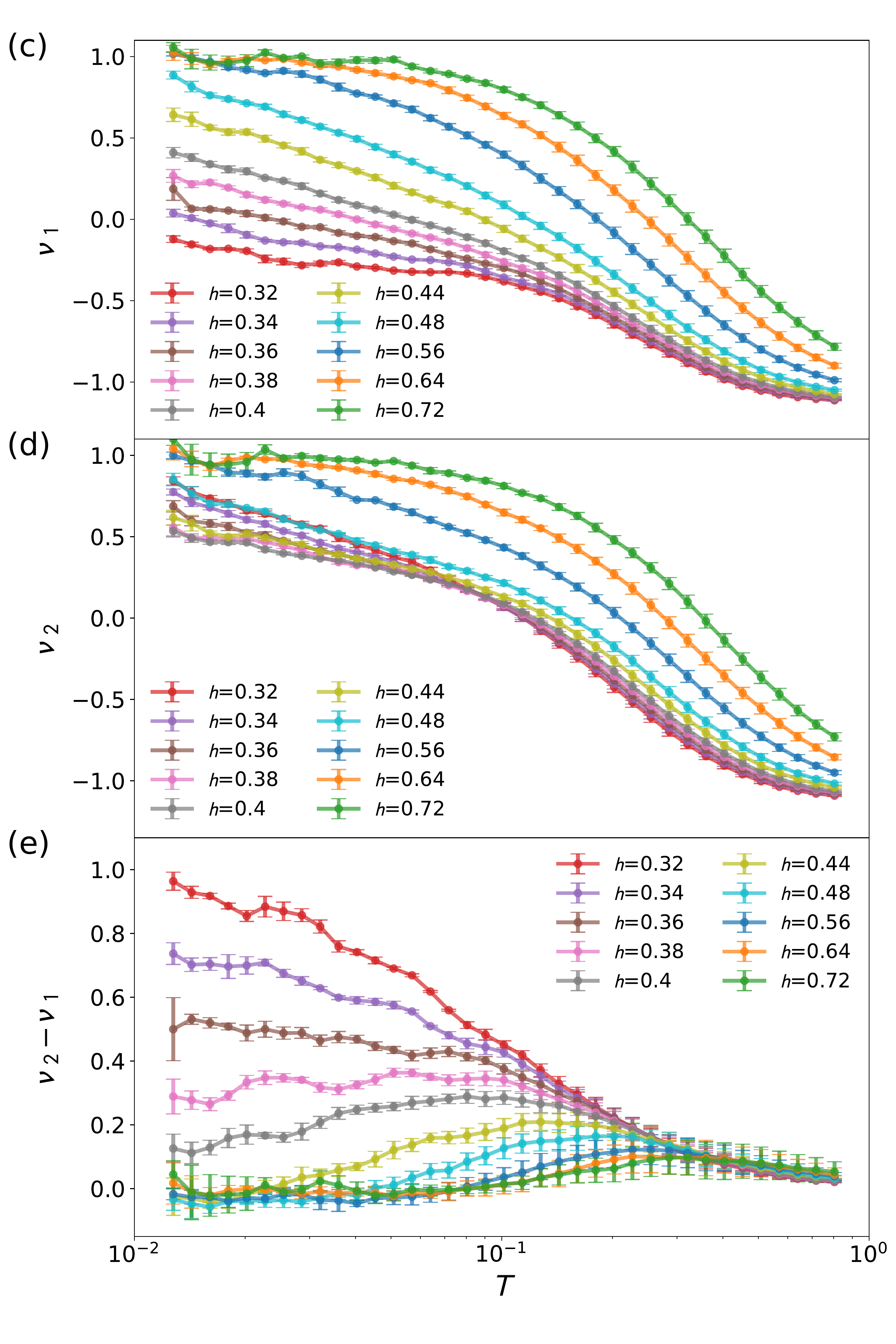}
        \caption{
Temperature dependences of Im$\Sigma_{\alpha}( i\omega_0 \rm{)}$ for the channel (a) $\alpha=1$ and (b) $\alpha=2$,  
and the exponent $\nu_\alpha$ [Eq.~(\ref{eq:nu})] for (c) $\alpha=1$ 
(d) $\alpha=2$, and (e) $\nu_2-\nu_1$ for several $h$.
        }
        \label{Fig:sigma}
      \end{minipage}
\end{tabular}
\end{flushleft}
\end{figure}

In order to understand the origin of the enhancement of $C/T$ in the PM phase, 
we analyze the imaginary part of the self-energy 
which represents the scattering of conduction electrons 
by the local moments, following the procedure in Sec.~\ref{subsec:thephysicalquantities}. 
The temperature dependences of ${\rm Im}\Sigma_{\alpha}(i\omega_0)$ are plotted for several $h$ in 
Figs.~\ref{Fig:sigma}(a) and \ref{Fig:sigma}(b) for the channel $\alpha=1$ and $2$, respectively. 
In the high-$h$ region, ${\rm Im}\Sigma_{\alpha}(i\omega_0)$ decreases while decreasing temperature, 
but turns to increase and approaches zero at low temperature in both channels. 
This is because $h_{\rm CEF}$ in Eq.~(\ref{eq:H_CEF}) aligns the local moments in parallel and 
reduces the scattering at low temperature; the minimum temperature of ${\rm Im}\Sigma_\alpha(i\omega_0)$ roughly corresponds to the energy scale of the internal magnetic field for conduction electrons, $J h_{\rm CEF}$. 
While decreasing $h$, Im$\Sigma_{\alpha}(i\omega_0)$ gradually grows to have larger negative values 
in both channels down to $h\sim0.40$. 
For $h\lesssim 0.40 $, however, Im$\Sigma_{\alpha}(i\omega_0)$ exhibits contrasting temperature dependence between the two channels: 
Im$\Sigma_{1}(i\omega_0)$ continues to decrease down to the lowest temperature calculated here,
while Im$\Sigma_{2}(i\omega_0)$ shows an upturn similarly to those for larger $h$. 
This differentiation of Im$\Sigma_{\alpha}(i\omega_0)$ indicates 
a large difference between the two channels with respect to
the scattering from the local moments. 

We estimate the power $\nu_\alpha$ defined in Eq.~(\ref{eq:nu}) for each channel by 
fitting the results in Figs.~\ref{Fig:sigma}(a) and \ref{Fig:sigma}(b). 
The fitting is done for five adjacent $T$ points including the focused temperature. 
The results are plotted in Figs.~\ref{Fig:sigma}(c) and \ref{Fig:sigma}(d) for $\alpha=1$ and $2$, respectively. 
In the  high-$h$ region, $\nu_\alpha$ 
approaches $1$ in both channels at low temperature, indicating that the FL state is realized.
While decreasing $h$, $\nu_\alpha$ is suppressed in both channels, 
but for $h\lesssim 0.40$, 
$\nu_1$ and $\nu_2$ behave differently;
$\nu_2$ turns to increase as decreasing $h$ and approaches $\sim 1$ at low temperature, 
while $\nu_1$ is further suppressed 
to $\sim 0$ at the lowest temperature calculated here. 
To clearly show the differentiation, we plot $\nu_2-\nu_1$ in Fig.~\ref{Fig:sigma}(e). 
This suggests that the channel $1$ behaves as a NFL state, 
while the channel $2$ is FL. 
We call this peculiar differentiation the CS-NFL behavior.

To clarify this peculiar behavior more explicitly, we summarize the estimates of 
$\nu_\alpha$ on the $h$-$T$ plane in Figs.~\ref{Fig:pd4}(c) 
and \ref{Fig:pd4}(d) for $\alpha=1$ and $2$, respectively.
The FL state where $\nu_\alpha \sim 1$ for both $\alpha=1$ and $2$ is extended in the high-$h$ PM region at low temperature. 
On the other hand, near the phase boundary of the AF-spin ordered phase, there is a window for $0.3\lesssim h \lesssim 0.4$ where the
CS-NFL behavior appears with $\nu_1 \sim 0$ and $\nu_2 \sim 1$.
In this region, 
$\Psi$ and $C/T$ are enhanced as plotted in Figs.~\ref{Fig:pd4}(a) and \ref{Fig:pd4}(b), respectively. 
Thus, our results indicate that the scattering from the local moments is differentiated in the region 
where $\Psi$ becomes large, and it leads to the CS-NFL state with the HF behavior in $C/T$.

The differentiation between $\nu_1$ and $\nu_2$ is found 
not only in the PM region but also inside the 
AF-spin ordered phase, as shown in Figs.~\ref{Fig:pd4}(c) and \ref{Fig:pd4}(d). 
The result indicates that the CS-NFL behavior 
appears in coincidence with the dome-shaped region 
where $\Psi$ is enhanced in Fig.~\ref{Fig:pd4}(a). 
This is reasonable because the scattering from the local moments 
can be channel selective when the composite correlation grows, irrespective of the AF-spin ordering. 

We note that, in the region of $0.4\lesssim h\lesssim 0.5$ between the CS-NFL and FL, 
both two channels appear to behave as NFL with $0<\nu_\alpha<1$. 
This NFL region roughly corresponds to the dip in $C/T$ found in Fig.~\ref{Fig:ct_phi_h}(d). 
We also note that $\nu_1$ becomes negative near $h\sim 0.10$, 
suggesting that the channel 1 is insulating while the channel 2 remains metallic. 
In this field region, we find that the electron filling in the channel 1 is almost fixed at half filling. 
Thus, it is regarded as the channel-selective Kondo insulating state. 
This is the regime where the suppression of $m_{\rm AF}$ and another enhancement of $C/T$ are 
observed in the previous sections. 
We note that the field range of the channel-selective Kondo insulator depends on the electron filling of the system.

\subsection{$J$ dependence}
\label{subsec:jdependences}

\begin{figure}[htbp]
\begin{tabular}{cc}
      \begin{minipage}[t]{1.0\hsize}
        \centering
        \includegraphics[width=8.5cm,pagebox=cropbox,clip]{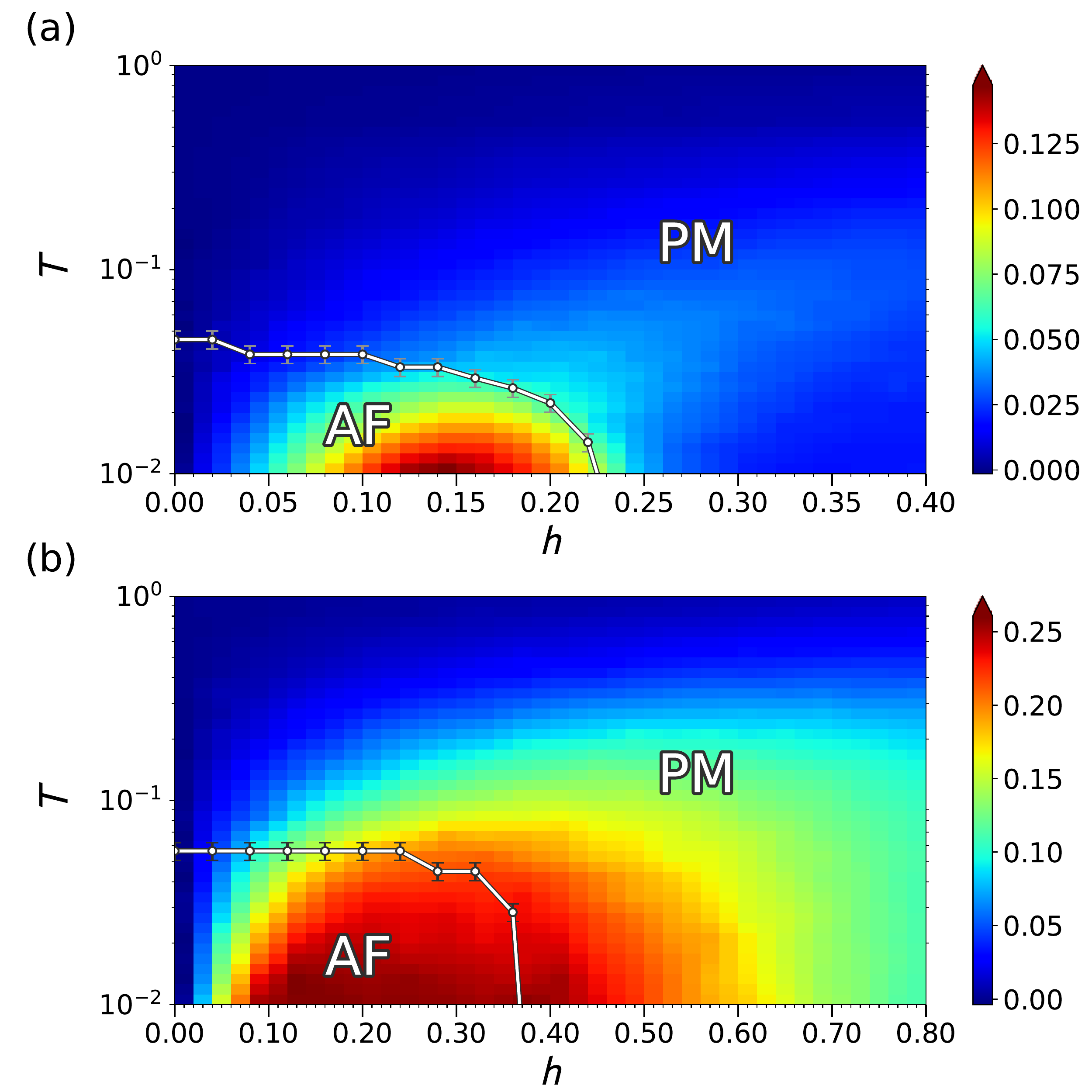}
        \caption{
Contour plots of $\Psi$ at (a) $J=0.4$ and (b) $J=1.2$. 
The white lines indicate the phase boundaries for the 
AF-spin ordered phase determined by the onset temperature of $m_{\rm AF}$.
}
        \label{Fig:pd_jdep}
      \end{minipage}
\end{tabular}
\end{figure}

We study the $J$ dependence of the CS-NFL state with enhanced $\Psi$. 
Figure~\ref{Fig:pd_jdep}(a) shows the phase diagram with the contour plot of $\Psi$ at a smaller $J=0.4$. 
In this case, the AF-spin ordered phase shrinks, 
and at the same time, the dome-shaped region where $\Psi$ is enhanced also shrinks; 
consequently, the dome is mostly contained within the AF-spin ordered phase. 
The result suggests that the HF behavior associated with the CS-NFL is hardly seen in the PM phase
outside the AF-spin ordered phase. 
On the other hand, as plotted in Fig.~\ref{Fig:pd_jdep}(b),
both the AF-spin ordered state and the dome-shaped region of $\Psi$ are extended for a larger $J=1.2$, and notably, 
the latter region is significantly extended to the PM state in a relatively wider region 
compared to the case with $J=0.8 $ in Fig.~\ref{Fig:pd4}. 
In this case, the HF behavior appears in the wider region (not shown). 
These results indicate that the magnitude of $J$ plays an important role in the parameter range of the CS-NFL and the HF behavior. 

It is worth noting that the AF-spin ordered phase continues to extend while increasing $J$ near half filling. 
This is in contrast to the ordinary Kondo lattice model where the AF phase is taken by a paramagnetic state for sufficiently large $J$ because of the Kondo singlet formation.

\section{Discussion}
\label{sec:discussion}

\subsection{Schematic phase diagram and comparison to the ordinary Kondo lattice model}
\label{subsec:magneticphasediagramoftcklmodel}

\begin{figure}[htbp]
\includegraphics[clip,width=7.5cm]{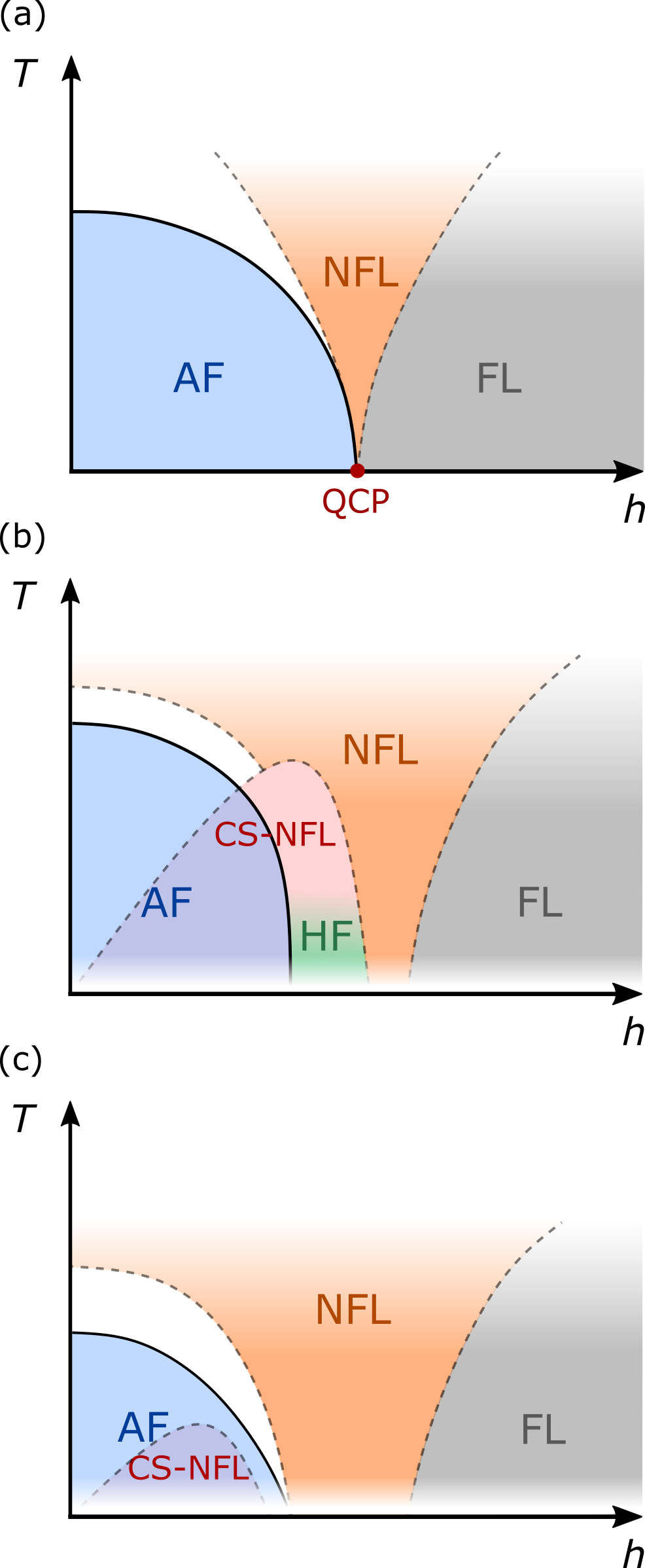}
\caption{
Schematic phase diagrams as functions of the magnetic field $h$ and temperature $T$ for 
(a) the ordinary Kondo lattice model, (b) the TCKL model with large $J$ 
and (c) small $J$. 
The solid and dashed lines indicate the phase boundaries and crossovers.
See the text for details. 
}
\label{fig:TCKL_PD}
\end{figure}

Bringing the obtained results together, we draw schematic phase diagrams for the TCKL model 
in a magnetic field for large and small $J$ 
in Figs.~\ref{fig:TCKL_PD}(b) and \ref{fig:TCKL_PD}(c), respectively. 
For comparison, we also show a schematic phase diagram for the ordinary Kondo lattice model 
in Fig.~\ref{fig:TCKL_PD}(a). 
In the case of the Kondo lattice model, 
the phase diagram is a variant of the Doniach phase diagram introduced in Sec.~\ref{introduction}~\cite{10.1038/nature03279,RevModPhys.79.1015,PhysRevLett.92.026401}, 
where the magnetically ordered phase competes with the PM phase. 
When the magnetic phase transition remains continuous down to zero temperature, 
NFL behavior appears in the quantum critical region near the QCP. 
In contrast, the phase diagram for the TCKL model becomes more complicated owing to the existence of both channel and spin degrees of freedom. 
The primary difference lies in the appearance of crossover by 
enhancement of the composite correlation in the dome-shaped region.
This crossover defines the region where the system exhibits the CS-NFL behavior. 
The CS-NFL region depends on the magnitude of $J$: 
It extends from the AF-spin ordered phase to the PM phase when $J$ is large [Fig.~\ref{fig:TCKL_PD}(b)], while it is 
included inside the AF-spin ordered phase when $J$ is small [Fig.~\ref{fig:TCKL_PD}(c)]. 
When $J$ is large, in addition, in the CS-NFL region on the PM side, the system shows HF behavior with enhanced $C/T$ at low temperature.
Meanwhile, in the PM state outside the CS-NFL region, the NFL behavior is observed 
in both channels before entering the higher-$h$ FL region as discussed in Sec.~\ref{subsec:channelselectivenonfermiliquidbehavior};
the NFL state is expected to extend in a wide region of $T$ and $h$ because it is not rooted 
in the QCP but in the overscreening nature inherent in the two-channel systems. 
Thus, the TCKL model exhibits much richer field-induced behavior than the Kondo lattice model.

We note that the phase diagram for the TCKL model also varies 
depending on the relation of the magnetic fields in Eq.~(\ref{eq:h}). 
Since the Zeeman field $h$ splits the energy levels of
the two channels, the CS-NFL dome is shrunk (extended) while increasing (decreasing) the coefficient in Eq.~(\ref{eq:h}) 
from the current value $\frac74$. 
When the coefficient is decreased sufficiently, the phase diagram is changed qualitatively, 
as an additional phase with a composite order intervenes between the AF-spin ordered and PM phases; see Fig.~\ref{fig:hhb100} in Appendix~\ref{app:hh} 
[note that the composite order parameter is different from $\Psi$ in Eq.~(\ref{eq:psi})]. 
Such a more complicated case is out of scope of the present study, 
as such successive transitions have not been 
observed yet in experiments.

\subsection{Implication to the 1-2-20 systems}
\label{subsec:therelationtothe1220systems}

Finally, let us discuss the possible implications of our results to the Pr-based 1-2-20 systems. 
As described in Sec.~\ref{introduction}, HF behavior was observed 
by the enhancement of $C/T$ 
as well as $T^2$ coefficient of the electrical resistivity 
in $\rm{PrIr_2Zn_{20}}$ and $\rm{PrRh_2Zn_{20}}$ under a magnetic field~\cite{PhysRevB.94.075134,doi:10.7566/JPSJ.86.044711}. 
The interesting point is that the HF behavior is observed not only in the vicinity of the QCP where the AFQ order disappears 
but also in a certain range of the field and temperature between the AFQ and FL states. 
Similar behavior is obtained in our results for the TCKL model: 
$C/T$ is enhanced in the CS-NFL region in the PM state 
where the composite correlation is enhanced. 
Hence, our finding provides a scenario that the HF behavior in these materials is caused by the NFL state in one of the spin components of conduction electrons, 
say spin-up, under the magnetic field (note that the channel in the TCKL model corresponds to spin in real materials). 
As mentioned above, the electrical resistivity in experiment shows $T^2$ behavior with enhanced coefficient in this region,
which is compatible with FL. 
It is left for future study to clarify how the resistivity behaves in the CS-NFL region in the TCKL model.
In addition, anomalous enhancement of the Seebeck coefficient 
was observed in the slightly higher-field region in experiments~\cite{doi:10.7566/JPSCP.3.011091,doi:10.7566/JPSJ.86.044711}. 
Since the region appears to correspond to the NFL region between the CS-NFL and FL states in our results, 
a possible scenario is that fluctuations from the NFL in both channels contribute to the enhancement of the Seebeck coefficient. 
The calculation of the Seebeck coefficient is also left for future study.

In $\rm{PrV_2Al_{20}}$, the AFQ phase extends in a wider range of temperature and field compared to PrIr$_2$Zn$_{20}$ 
and PrRh$_2$Zn$_{20}$~\cite{PhysRevB.91.241102}. 
Correspondingly, NFL behavior peculiar to the quadrupole Kondo systems 
was observed, e.g., in the $\sqrt{T}$ scaling of electrical resistivity, 
in a much wider region below $30$~K~\cite{doi:10.7566/JPSJ.89.013704}. 
Below $8$~K, however, this scaling no longer holds and the system exhibits power-law divergence in the specific heat~\cite{doi:10.7566/JPSJ.89.013704}.
This suggests a crossover from NFL to HF states while decreasing temperature. 
These behaviors are at least qualitatively consistent with our results for the TCKL model for large $J$. 
Indeed, a large $c$-$f$ coupling has been pointed out for $\rm{PrV_2Al_{20}}$ 
compared to PrIr$_2$Zn$_{20}$ and PrRh$_2$Zn$_{20}$~\cite{PhysRevLett.113.267001,1742-6596-592-1-012025,PhysRevB.88.085124}.

These comparisons suggest the possibility that the peculiar behaviors in the 1-2-20 compounds, 
NFL, FL, and HF states and crossover between them, can be qualitatively understood by the TCKL model including the two types of the magnetic fields. 
In particular, our scenario proposes that there are two types of the NFL regimes, and the HF behavior is 
associated with one of them, namely, the CS-NFL state in the PM region (spin-selective NFL in reality). 
Such a scenario would be tested by systematically investigating the compounds with different magnitude of $J$ ($c$-$f$ coupling) 
and also by applying pressure to control the bandwidth and $J$.

\section{Summary}
\label{sec:summary}

In summary, we have clarified unconventional behaviors in the TCKL model in the magnetic field. 
To reproduce a realistic situation in quadrupole Kondo systems, 
we incorporated two different types of the magnetic fields: 
the Zeeman splitting for the conduction electrons and the CEF splitting 
for the localized moments. 
By using the CDMFT combined with the CTQMC method, 
we unveiled that the model exhibits the CS-NFL state 
in which only one of the channels behaves as a NFL while the other remains as a FL.
This peculiar state appears in the dome-shaped region extending from the AF-spin ordered state to the PM state, where the composite correlation 
between the conduction electron spin and localized moment is enhanced by the magnetic field through the imbalance between the two channels. 
Furthermore, we found HF behavior with an increase 
of $C/T$ in the CS-NFL state protruding to the PM side.
Thus, the HF behavior in our TCKL model is observed in a certain region of the field and temperature near the AF-spin ordered state. 
These behaviors are in stark contrast to those in the ordinary Kondo lattice model where 
the NFL behavior is limited to a narrow quantum critical region. 
We also showed that the extent of the CS-NFL region depends on the value of $J$ as well as the relative magnitude of the two types of the magnetic fields. 
We discussed that our findings of the CS-NFL and HF behaviors may provide a unified understanding of the experimental results in the 1-2-20 compounds such as 
$\rm{PrIr_2Zn_{20}}$, $\rm{PrRh_2Zn_{20}}$, and $\rm{PrV_2Al_{20}}$.

While our results have unveiled interesting properties of the TCKL model, there remain several issues to be clarified. 
One is the effect of anisotropy in the quadrupole degree of freedom. 
In our model, we assume the isotropic coupling between the conduction electron spins and the localized moments, but in reality, the coupling could be anisotropic reflecting the crystal symmetry. 
Indeed, the 1-2-20 compounds exhibit different behavior depending on the direction of the magnetic field~\cite{doi:10.1143/JPSJ.80.093601,PhysRevB.86.184426,PhysRevB.91.241102, doi:10.7566/JPSJ.82.043705}. 
It would be interesting to study the effect of anisotropy by extending our model. 
Another interesting issue is the possibility of superconductivity. 
In some 1-2-20 compounds, superconductivity is found inside the AFQ phase
~\cite{PhysRevLett.106.177001,PhysRevLett.113.267001,PhysRevB.86.184426}. 
While the superconductivity was studied for the TCKL model by the DMFT~\cite{PhysRevLett.112.167204},
it would be intriguing to investigate this issue by a straightforward extension of our CDMFT which can treat the competition between superconductivity and the AF-spin order on an equal footing.

\begin{acknowledgments}
The authors thank J. Yoshitake for constructive comments on the CDMFT and CTQMC methods. 
They also thank S. Hoshino, K. Izawa, H. Kusunose, and K. Matsubayashi for fruitful discussions.
The numerical calculations were performed on the facilities of 
the Supercomputer Center, the Institute for Solid State Physics, the University of Tokyo.
One of the authors, (K. I.) was supported by the Japan Society for the Promotion of Science through Program for Leading Graduate Schools (MERIT).
\end{acknowledgments}

\appendix

\section{Effects of two types of magnetic fields}
\label{app:hh}

\begin{figure}[htbp]
  \centering
  \begin{tabular}{cc}
      \begin{minipage}[t]{1.0\hsize}
        \includegraphics[clip,width=8.0cm]{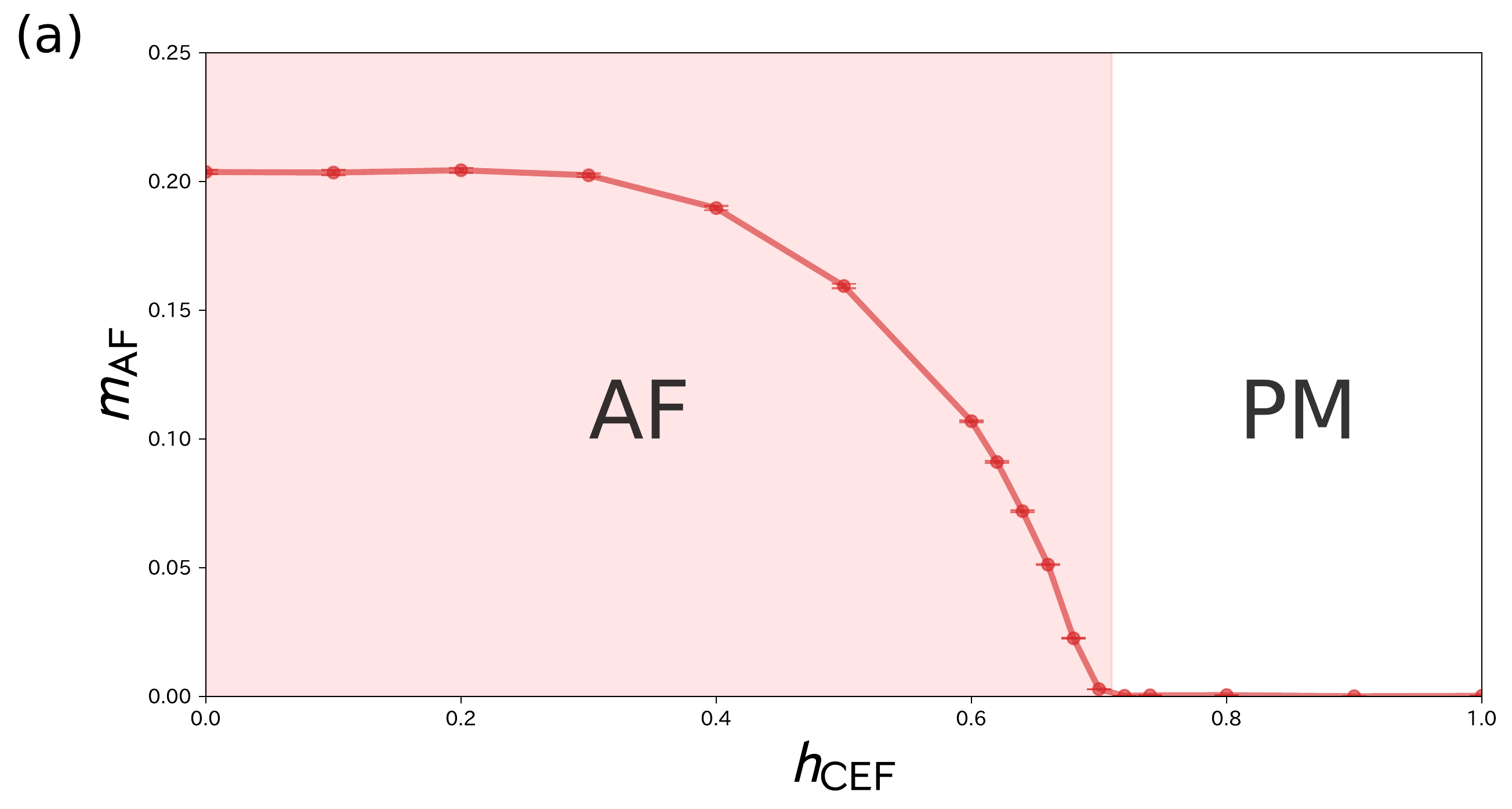}
        \label{fig:d9z0cd}
      \end{minipage}
      \\
      \begin{minipage}[t]{1.0\hsize}
        \includegraphics[clip,width=8.0cm]{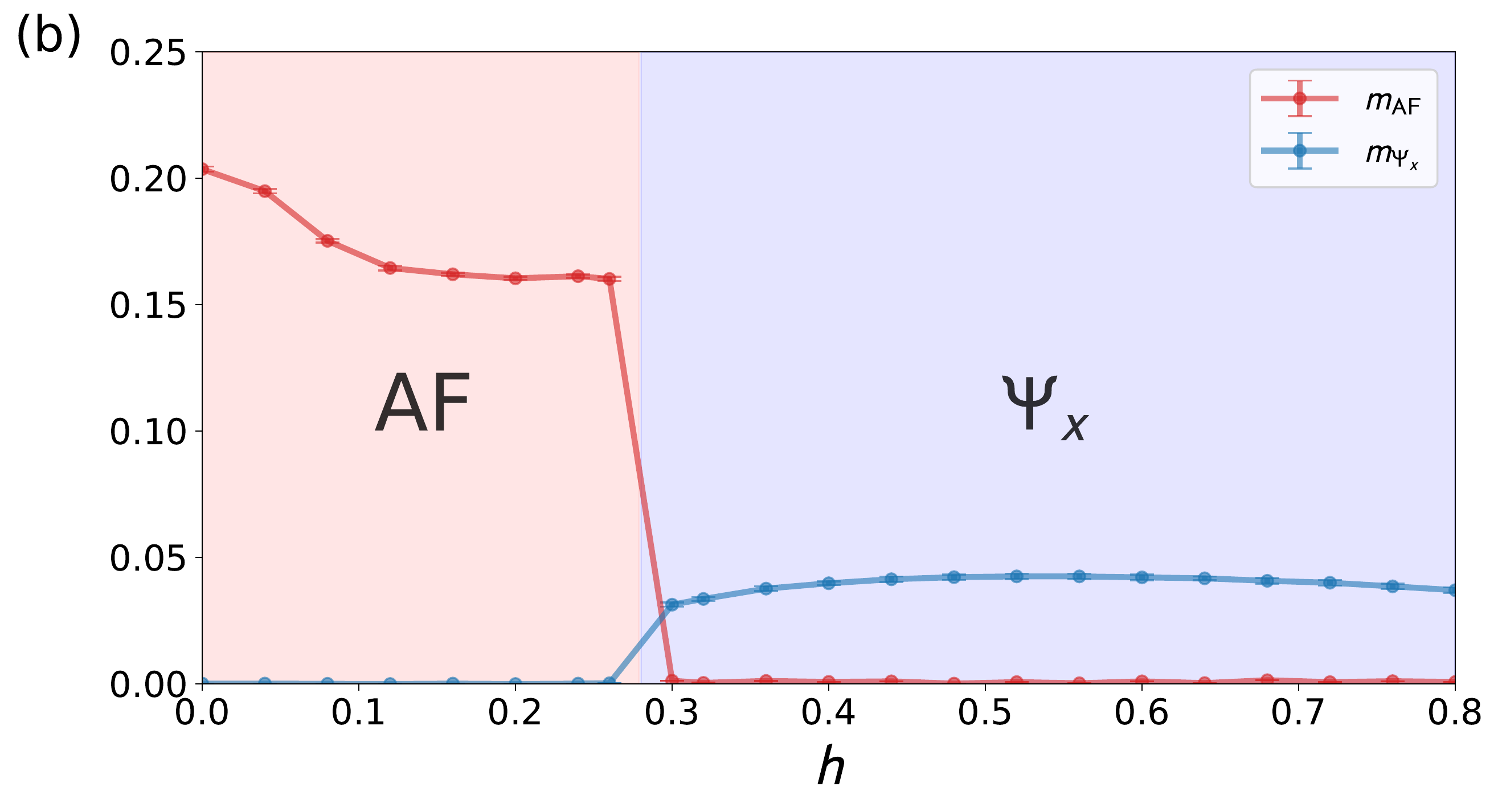}
        \label{fig:d9zdc0}
      \end{minipage} 
  \end{tabular}
  \caption{
      (a) $h_{\rm CEF}$ dependence of $m_{\rm AF}$ in Eq.~({\ref{eq:af}}) at $h=0$ and 
      (b) $h$ dependence of $m_{\rm AF}$ and $m_{\rm \Psi_x}$ in Eq.~(\ref{eq:psix}) at $h_{\rm CEF}=0$. 
      The red (blue) hatched region represents the AF-spin ($\Psi_x$) ordered phase, while the white region is the PM phase.
      We take $J=0.8$, $T=0.01$, and $n=0.9$. 
     }
  \label{fig:d9}
\end{figure}

In this Appendix, we discuss the effects of the two types of magnetic fields, 
$h$ and $h_{\rm CEF}$ in Eqs.~(\ref{eq:H_Zeeman}) and (\ref{eq:H_CEF}), respectively.
In the main text, the two fields are applied simultaneously 
with the relation in Eq.~(\ref{eq:h}), but here we study their effects independently.

Figure~\ref{fig:d9}(a) shows the AF-spin order parameter $m_{\rm AF}$ 
as a function of $h_{\rm CEF}$. 
We set $h=0$ with $J=0.8$, $T=0.01$, and $n=0.9$. 
We find that $m_{\rm AF}$ is gradually suppressed by $h_{\rm CEF}$ and continuously goes to zero, 
which indicates a second-order phase transition from the AF-spin ordered phase to the PM phase. 
On the other hand, $h$ leads to qualitatively different 
behavior as shown in Fig.~\ref{fig:d9}(b). 
Here we set $h_{\rm CEF}=0$. 
In this case, $m_{\rm AF}$ disappears abruptly at $h\simeq 0.3$. 
For larger $h$, we find that another order parameter becomes nonzero that is defined by 
\begin{eqnarray}
m_{\Psi_x} = \frac{1}{2}\sum_{\langle i,j \rangle} \sum_{
\alpha \alpha^\prime \sigma} \langle c_{i\alpha \sigma}^{\dagger}\sigma_{\alpha \alpha^\prime}^x c_{j\alpha^\prime \sigma} \rangle. 
\label{eq:psix}
\end{eqnarray}
Note that this is the $x$ component of $\bm{\Psi}_c(\bm{0})$ 
in Eq.~(15) in Ref.~\cite{PhysRevLett.112.167204}, which corresponds to a different type of the composite correlation from Eq.~(\ref{eq:psi}). 

\begin{figure}[H]
\includegraphics[clip,width=8.0cm]{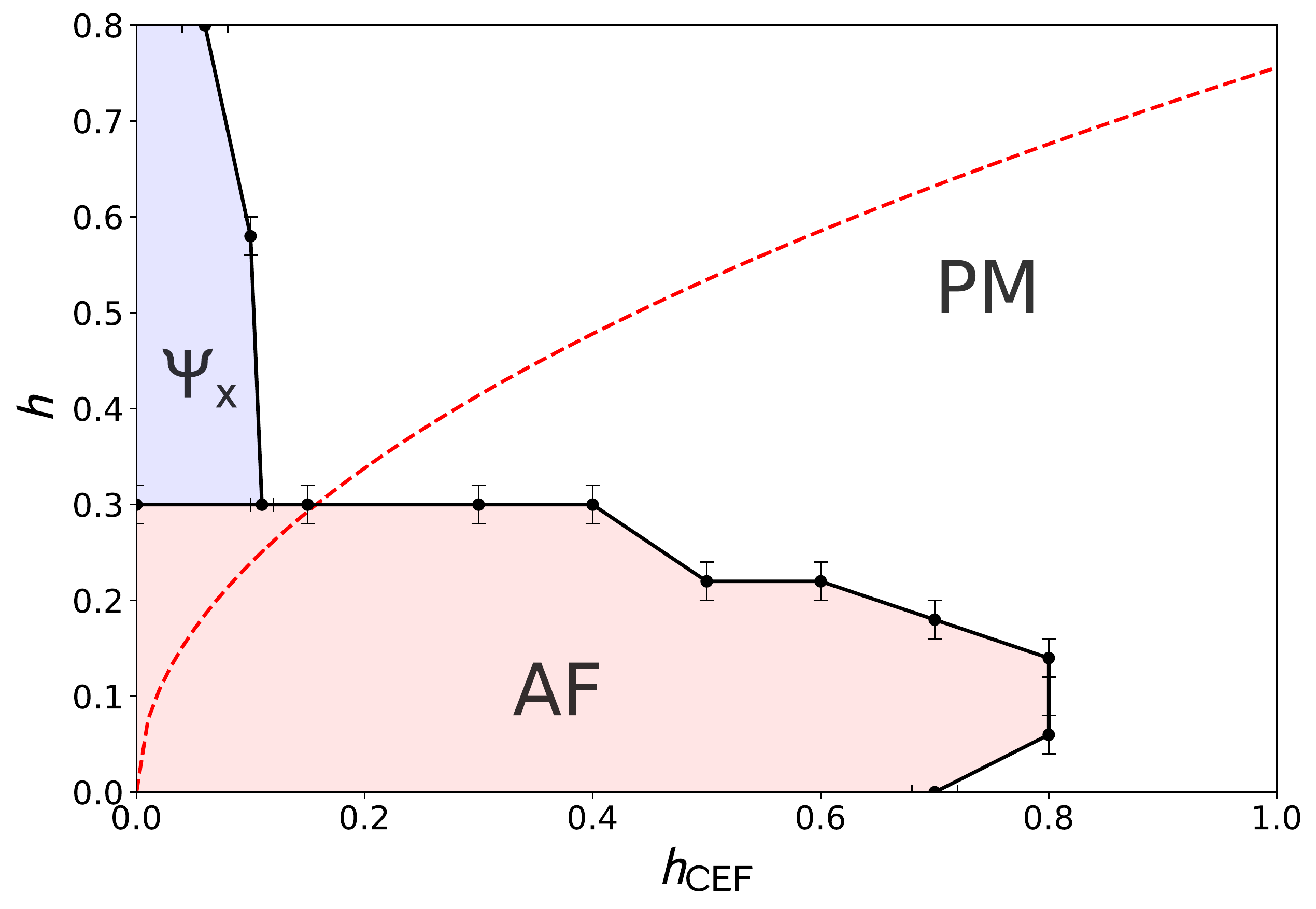}
\caption{
Phase diagram of the model in Eq.~(\ref{eq:H_total}) as a function of $h$ and $h_{\rm CEF}$ at $J=0.8$, $T=0.01$, and $n=0.9$. 
The red dotted curve follows the relation in Eq.~(\ref{eq:h}). 
}
\label{fig:hhb100}
\end{figure}

Figure~\ref{fig:hhb100} shows the phase diagram in the plane of $h_{\rm CEF}$ and $h$ at $T=0.01$. 
We take $J=0.8$ and $n=0.9$. 
The AF-spin ordered phase 
remains robust with spin canting up to $h_{\rm CEF}\simeq 0.8$. 
On the other hand, the $\Psi_x$ ordered phase is fragile against $h_{\rm CEF}$ since the 
singlet formation is destroyed by $h_{\rm CEF}$.
The red dotted curve corresponds to the relation in Eq.~(\ref{eq:h}) used for the calculations in the main text.
The result shows that when the coefficient in Eq.~(\ref{eq:h}) is decreased sufficiently, 
the $\Psi_x$ ordered phase intervenes between the AF-spin ordered and PM phases, as stated in Sec.~\ref{subsec:magneticphasediagramoftcklmodel}.

%\bibliography{main}% Produces the bibliography via BibTeX.

%merlin.mbs apsrev4-1.bst 2010-07-25 4.21a (PWD, AO, DPC) hacked
%Control: key (0)
%Control: author (72) initials jnrlst
%Control: editor formatted (1) identically to author
%Control: production of article title (-1) disabled
%Control: page (0) single
%Control: year (1) truncated
%Control: production of eprint (0) enabled
%

\end{document}